\documentstyle[11pt,appb,epsf]{article}

\def\lsim{\:\raisebox{-0.5ex}{$\stackrel{\textstyle<}{\sim}$}\:}

\begin{document}
\title{LOW X INELASTIC ELECTRON SCATTERING \\
FROM GENERALIZED VECTOR DOMINANCE\thanks{Invited talk presented at the 
    XXI School of Theoretical Physics, Ustro\'n, Poland, September 1997,
    to appear in Acta Physica Polonica.}
\thanks{Supported by the
    Bundesministerium f\"ur Bildung und Forschung, Bonn, Germany,
    Contract 05 7BI92P (9) and the EC-network contract CHRX-CT94-0579.}
}
\author{D.\ Schildknecht 
\address{Fakult\"at f\"ur Physik, Universit\"at Bielefeld, 
D-33501 Bielefeld}
}
\maketitle
\begin{abstract}
\centerline{{\bf Abstract}}

It is shown that the HERA experimental data on deep inelastic 
scattering at low values of the scaling variable $x \leq 0.05$ are 
in good agreement with predictions from Generalized Vector Dominance in the 
full kinematic range from $Q^2 = 0$ (photoproduction) to $Q^2 \simeq
350$ GeV.

\end{abstract}
\PACS{12.40.Vv, 13.60.-r, 13.60.Hb}

\bigskip
  

This morning we heard several talks on various aspects of deep inelastic 
scattering (DIS) of electrons on protons, in particular in the 
energy range presently explored by the H1 and ZEUS collaborations at HERA 
in Hamburg. In the talk by 
Kwiecinski \cite{1} theory and experiment were confronted at all 
values of the Bjorken scaling variable $0 < x_{bj} < 1$ {\it excluding} the 
$Q^2 = 0$ photoproduction limit. Spiesberger \cite{2} and  Kalinowski
\cite{3} concentrated on the anomalous events observed \cite{4} by the 
H1 and ZEUS collaborations at HERA at large values of $x \equiv x_{bj}$. 
In the present talk, essentially based on a recent paper \cite{5} by
Hubert Spiesberger and myself, I will examine the region of small
$x \simeq \frac{Q^2}{W^2} \ll 10^{-2}$, {\it including} 
the photoproduction, $Q^2 = 0$, limit. 

The motivations for this work are twofold:
\begin{itemize}
\item[i)]
At HERA two interesting experimental results at low $x$ were established
since HERA started operating in 1992: First of all, the proton
structure function $F_2 (x, Q^2)$ rises steeply with 
decreasing $x \leq 10^{-2}$ and shows a considerable amount of 
scaling violation \cite{5a}. Secondly, when analysing the final hadronic state 
produced, the H1 and ZEUS collaborations found an appreciable fraction of 
final states (approximately 10\% of the total) of 
typically diffractive nature (``large rapidity gap events'') with 
invariant masses of the produced hadronic state up to about 20 GeV \cite{6}. 
\item[ii)]
With respect to DIS at small $x$, a long-standing theoretical question 
concerns the 
role of the variables $x$ and $Q^2$. This question has been most succinctly
posed and discussed by Sakurai and Bjorken, as recorded in the
Proc. of the '71 Electron Photon Symposium at Cornell University \cite{7}. It 
concerns the transition to the hadronlike behaviour of photoproduction, in 
particular, whether a behaviour similar to photoproduction sets in in the 
limit of $Q^2 \rightarrow O$ only, or rather in the limit of 
$x\rightarrow 0$ at arbitrarily large fixed values of $Q^2$. Even after 
the formulation of QCD there is not yet a unique answer to this question. 
It is likely that the HERA low $x$ data in conjunction with 
theoretical analysis will resolve this important issue.
\end{itemize}

In the paper by Spiesberger and myself, we take the point of view that 
indeed $x$ is the relevant variable, in the sense that $x \leq 10^{-2}$
defines the region in which those features of the virtual 
photoproduction cross section, $\sigma_{\gamma^* p}$,
become important which show a close similarity to photoproduction and 
hadron-induced processes (Generalized Vector Dominance \cite{8}).

As a starting point, let me briefly remind you of photoproduction. 
``Ha\-dron\-like'' behaviour of photoproduction is quantified by relating
\cite{9} the high-energy forward Compton scattering amplitude on protons
to the forward amplitude of vector-meson proton scattering, and 
consequently also to vector-meson forward production extrapolated to 
$t = 0$ (where $t$ is the four-momentum transfer squared to the initial 
proton),
\begin{equation}
\sigma_{\gamma p} (W^2) = \sum_{\rho^0, \omega, \phi, J/\psi} 
\sqrt{16\pi} \sqrt{\frac{\alpha\pi}{\gamma^2_V}} \left( \frac{d\sigma}{dt}
\biggl|_{t = 0} \biggr. (W^2) \right)^{\frac{1}{2}}.
\label{(1)}
\end{equation}
The photon vector-meson couplings have to be extracted from $e^+ e^-$
annihilation, i.e.
\begin{equation}
\frac{\alpha\pi}{\gamma^2_V} = \frac{1}{4\pi^2 \alpha}\, \sum_F \,
\int \, \sigma_{e^+ e^- \rightarrow V \rightarrow F}
 (s) \, ds ,
\label{(2)}
\end{equation} 
where the integral is extended over the peak of the cross section 
corresponding to production of the vector meson $V$ with subsequent 
decay into the final state $F$. 
The sum rule (\ref{(1)}) may be rewritten as 
\begin{equation}
\sum_V r_V = 1
\label{(3)}
\end{equation}
with 
\begin{equation}
r_V = \frac{1}{\sigma_{\gamma p}} \sigma_{_{V p}} \frac{\alpha\pi}
{\gamma^2_V} =
\frac{1}{\sigma_{\gamma p}} \sqrt{16\pi} \sqrt{\frac{\alpha\pi}{\gamma^2_V}} 
\left( \frac{d\sigma}{dt} \biggl|_{t=0} \biggr. (\gamma p \rightarrow V p )
\right)^{\frac{1}{2}}.
\label{(4)}
\end{equation}
Experimentally, from $e^+ e^-$-annihilation and photoproduction, one finds that
$\rho^0, \omega$ and $\phi$ fail to saturate the sum rule at the level of 
22\%, as \cite{8}
\begin{equation}
\sum_{\rho^0, \omega , \phi} r_V = 0.78.
\label{(5)}
\end{equation}

Generalized Vector Dominance \cite{8} starts from the hypothesis that the 
22\% defficiency in the sum rule (\ref{(3)}) is made up by the contribution 
of the more massive states directly produced in $e^+ e^-$-annihilation.
Indeed, the propagation of these more massive states increases their 
weight at spacelike four-momenta of the (virtual) photon considerably, 
compared with the $\rho^0 , \omega , \phi$ contributions which are of minor 
importance, once $Q^2$ becomes large compared with the mass of the $\rho^0$
meson, $Q^2 \gg m^2_\rho$. 
Accordingly, in low $x$ DIS in Generalized Vector Dominance, one 
expects an appreciable signal for diffractive production of high-mass 
states. Indeed, the HERA experiments found such a signal \cite{6}.
Moreover, shadowing in inelastic scattering from complex nuclei, as a 
result of diffractive production of high-mass states,
was expected to persist at low $x$ \cite{10} and large $Q^2$, which is indeed
the case \cite{11} in semiquantitative agreement with the predictions 
\cite{10,12}. 
These features of DIS, persistance of shadowing for 
spacelike $Q^2$ and diffractive production of high-mass states, are 
suggestive of treating low-$x$ DIS quantitatively from the point of view of 
Generalized Vector Dominance. 

In the diagonal approximation, the transverse part of the 
photon absorption cross section, $\sigma_T (W^2, Q^2)$, reads \cite{8}
\begin{equation}
\sigma_T (W^2, Q^2 ) = \int_{m^2_0} \, dm^2 \, \frac{\rho_T (W^2 , m^2)
m^4}{(m^2 + Q^2 )^2} , 
\label{(6)}
\end{equation}
where the spectral weight-function, $\rho_T$, is proportional to the 
product of the cross
section of $e^+ e^-$ annihilation into hadrons at the energy $m^2$ and the 
hadronic cross section for the scattering of the state of mass $m^2$ on the 
nucleon, 
\begin{equation}
\rho_T (W^2 , m^2 ) = \frac{1}{4\pi^2 \alpha} \sigma_{e^+e^-} (m^2)
\sigma_{{\rm hadr}} (W^2 , m^2).
\label{(7)}
\end{equation}
The cross section $\sigma_{{\rm hadr}}$ is clearly to be identified with 
the total cross section for scattering on the nucleon of the state of mass 
$m$, which apart from being producible in $e^+e^-$ annihilation, should also 
be visible in diffractive production by (virtual) photons on protons 
(``large rapidity gap events''). 

In the recent paper \cite{5} we concentrated on evaluating (\ref{(6)}) 
in the high-energy limit, $W \geq 60$ GeV, where hadronic and photoproduction
cross sections rise with increasing energy. We adopted an ansatz with a 
logarithmic rise, 
\begin{equation}
\rho_T (W^2 , m^2 ) = N \frac{\ln (W^2 / a m^2)}{m^4} , 
\label{(8)}
\end{equation}
which obviously fails to describe experimental data a lower energy. 
The $m^{-4}$ dependence contains an $m^{-2}$ factor from $e^+ e^-$
annihilation and an $m^{-2}$ factor from the subsequent interaction of this 
state with the nucleon. 
This latter $m^{-2}$ factor has to be considered as a theoretical input 
which is enforced, if powerlike (in $Q^2$) scaling 
violations of the proton structure function $F_2$ are to be excluded. 
Bjorken argues \cite{7,13} that jet alignment may be the origin of the 
$m^{-2}$ decrease in the scattering of the state of mass $m \gg m_\rho$ from 
the nucleon, even though in $e^+ e^-$ annihilation the 2-jet $(q \bar q)$ 
configuations are not very 
pronounced in the mass range ($m < 20$ GeV) relevant at present HERA 
energies. The threshold mass $m_0 \approx m_\rho$ in 
(\ref{(6)}) in principle is determined by $e^+ e^-$ annihilation, while the 
normalization $N$ and the scale $a$ of the energy dependence are
determined by the $Q^2 = 0$ photoproduction limit of $\sigma_T$ in 
(\ref{(6)}) upon substitution of (\ref{(8)}). 
In the recent paper \cite{5}, these parameters were actually determined by a
fitting procedure which included $Q^2 \not = 0$ electron-scattering data
as well as $Q^2 = 0$ photoproduction.  

Before turning to the analysis of the experimental data, the 
extension of (\ref{(6)}) to production of hadrons by longitudinal 
photons, $\sigma_L (W^2 , Q^2)$, has to be given. Introducing the ratio $\xi$
of longitudinal-to-transverse (on-mass-shell) scattering of the state of mass
$m$, we have \cite{8}
\begin{equation}
\sigma_L (W^2 , Q^2 ) = \int_{m^2_0} \, dm^2 \frac{\xi \rho_T (W^2 , m^2) 
m^4}{(m^2 + Q^2)} \, \frac{Q^2}{m^2} \nonumber , 
\label{(9)}
\end{equation}
where the factor $Q^2/m^2$ originates from the coupling of the 
hadronic vector state of mass $m$ to a conserved source \cite{14}. 
The integration in (\ref{(6)}) and (\ref{(9)}) 
may be carried out in closed form. For the resulting expressions we refer
to the original paper \cite{5}.

>From the fit to the H1 and ZEUS data \cite{5a} we obtained
\begin{eqnarray}
N & = & 0.187 \times 4 \pi^2 \alpha = 0.054 , \nonumber \\ 
m^2_0 & = & 0.89 {\rm GeV}^2 , \\
\xi & = & 0.171 , \nonumber \\
a & = & 15.1 . \nonumber
\label{(10)}
\end{eqnarray}
The fact that the threshold mass $m_0$ is somewhat larger than the 
$\rho^0$ mass, $m_\rho$, is presumably due to our very simplified 
ansatz which 
does not discriminate between the different thresholds associated with the 
light-quark and the charm-quark masses. Figure 1 shows remarkably good 
agreement for 
\begin{figure}[htb] 
\unitlength 1mm
\begin{picture}(125,150)
\put(-30,-42){
\epsfxsize=17.2cm
\epsfysize=24cm
\epsfbox{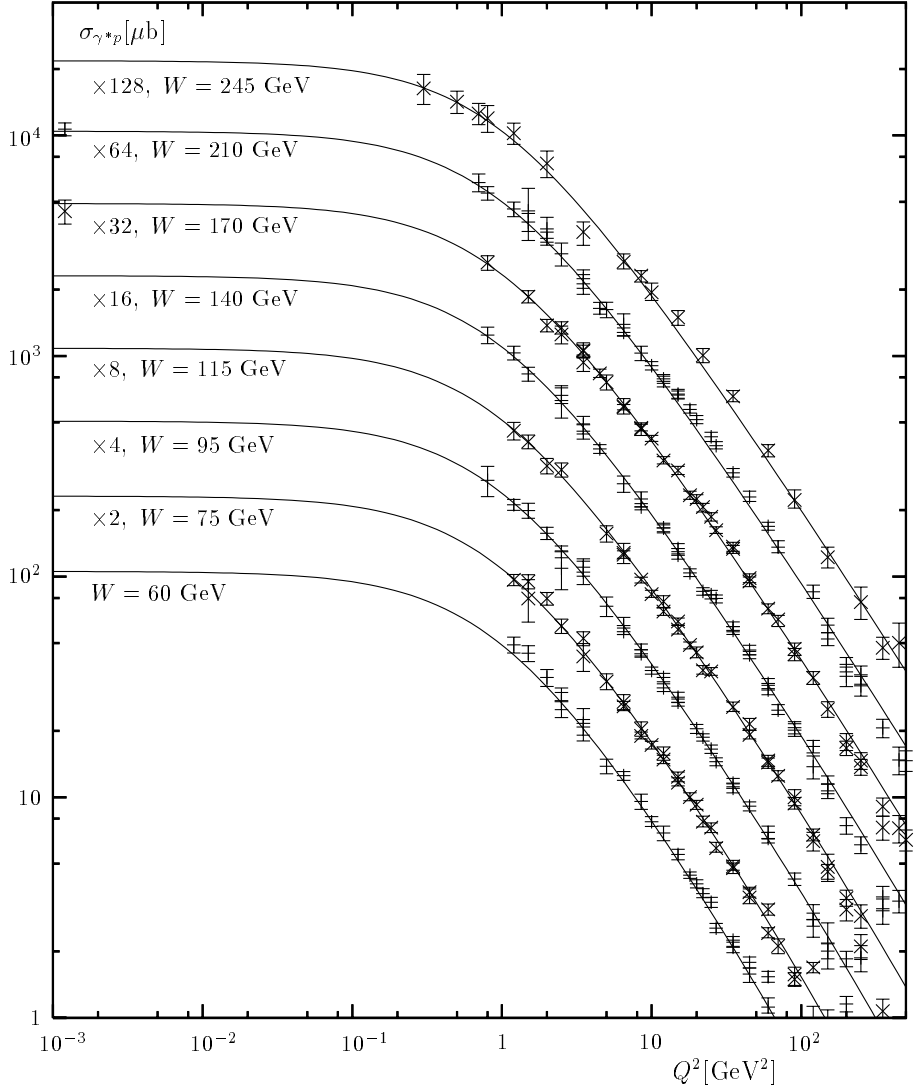}}
\end{picture}
\caption{\it 
  Generalized Vector Dominance prediction for $\sigma_{\gamma^*p}$ 
  compared with the experimental data from the H1 and ZEUS
  collaborations at HERA.}
\label{fig2}
\end{figure}
\begin{equation}
\sigma_{\gamma * p} \equiv \sigma_T + \sigma_L  
\label{(11)}
\end{equation}
with the experimental data over the full $Q^2$ range from photoproduction,
$Q^2 = 0$,
to $Q^2 \simeq 350 {\rm GeV}^2$ at energies from $W \simeq 60$ GeV to 
$W\simeq 245$ GeV, corresponding to values of the scaling variable 
$x \leq 0.05$.

Figure 2
shows the proton structure function
\begin{equation}
F_2 (W^2 , Q^2 ) \simeq \frac{Q^2}{4\pi^2 \alpha} (\sigma_T + \sigma_L )
\label{(12)}
\end{equation}
as a function of $x \simeq Q^2/W^2$ for various values of $Q^2$.
This Figure explicitly shows that the theoretical prediction for the 
transverse part of $F_2$, due to $\sigma_T$, in our simple ansatz has 
reached its scaling limit for $Q^2 > 12{\rm GeV}^2$. 
The rise of $F_2 (W^2 , Q^2)$ with increasing $Q^2$ for 
$Q^2 > 12 {\rm GeV}^2$ is due to the influence of $\sigma_L$. 
For details we refer to the original publication \cite{5}. 

\begin{figure}[htbp] 
\unitlength 1mm
\begin{picture}(125,150)
\put(-19,-35){
\epsfxsize=15cm
\epsfysize=21cm
\epsfbox{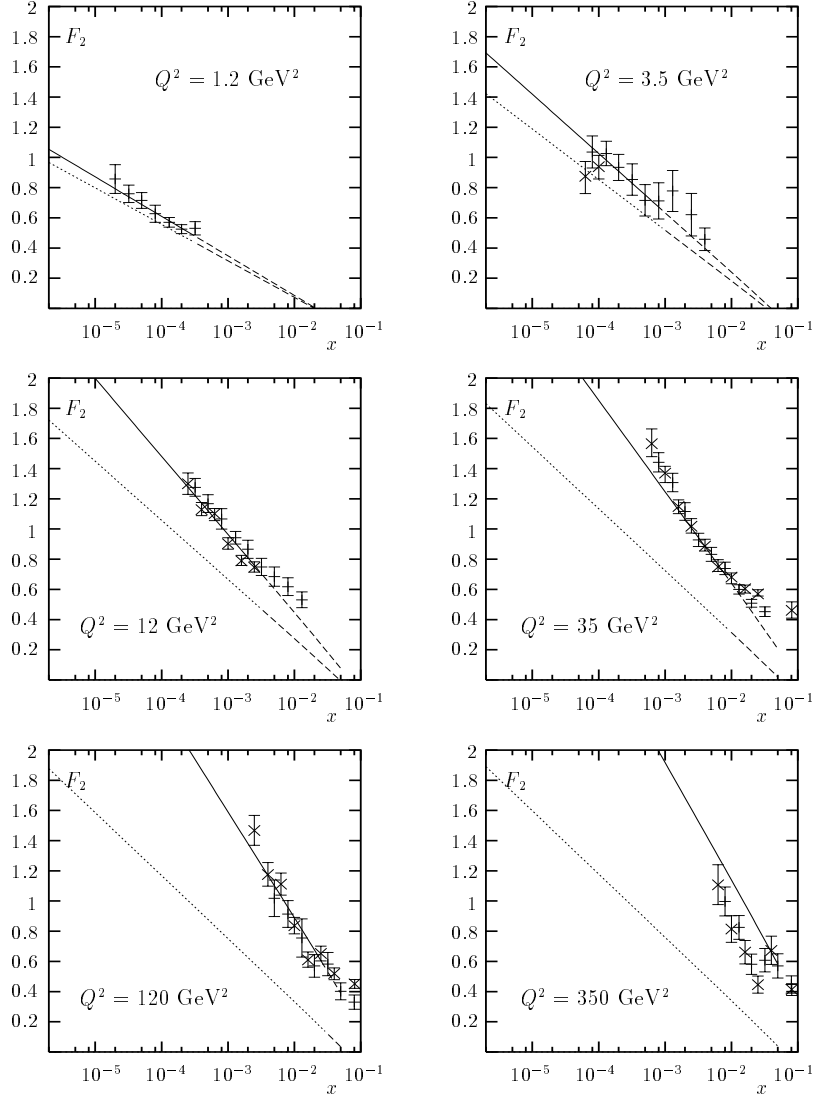}}
\end{picture}
\caption{\it
  Generalized Vector Dominance prediction for $F_2$ as a function of $x$
  compared with the HERA data for different values of $Q^2$. The dotted
  lines show the contribution to $F_2$ due to transverse virtual photons
  $(\xi = 0)$. The dashed curves indicate the
  region $W \protect\lsim 60$ GeV in which the present model becomes
  inadequate.}
\label{fig4}
\end{figure}

Various refinements of the present work are to be carried out, such as the 
extension of this high-energy model to lower values of $W$, the 
incorporation of the charm threshold, an analysis of how additional scaling 
violations in $\sigma_T$ may appear within the present 
non-perturbative ansatz, etc. 
Further experimental tests consist of separating $\sigma_L$ and $\sigma_T$ 
experimentally, a project for the more distant future, while more detailed
experimental studies of diffractive production in the nearby future 
will provide tests of the 
underlying theoretical assumptions. 

Various parametrizations of the experimental data on low $x$ deep 
inelastic scattering, including photoproduction, exist in the literature,
either based on \cite{15,16,17} 
modifications of Regge theory or on a combination \cite{18}
of $\rho^0, \omega, \phi$ dominance with the parton-model approach. The 
fit of the data presented in \cite{19}
is of interest in the context of the present paper, as logarithmic $Q^2$
and $x$ dependences only are employed in the fit. An analysis of the data
which in its spirit is similar to the one of the present paper, even though 
different in detail, is given in \cite{20}.

In summary, it has been shown that the framework of Generalized Vector 
Dominance is able to provide a unified representation of 
photoproduction and the low-$x$ proton structure function 
in the kinematic range accessible to HERA which is in good agreement 
with experiment. Details are subject to improvement and change in close 
collaboration between theory and experiment. The principal dynamical 
ansatz, relating $\sigma_{\gamma * p}$, or , equivalently, $F_2$  at 
low values of $x$ to 
diffractive scattering (via unitarity) of the states produced in $e^+ e^-$
annihilation, is likely to stand the test of time.

\bigskip
\noindent{\large \bf Acknowledgment}\\[1mm] 
It is a pleasure to thank Hubert Spiesberger for fruitful collaboration on 
the subject matter of the present report and Karol Kolodziej and his 
organizing committee for the 
invitation to the lively and successful physics meeting in Ustro\'n.

\bigskip

\end{document}